\documentclass[pra, aps, twocolumn, groupedaddress, superscriptaddress]{revtex4}

\usepackage{slashed}
\usepackage{graphicx}
\usepackage{subfigure}
\usepackage[usenames, dvipsnames]{color} 
\usepackage{graphics}
\usepackage{hyperref}
\usepackage{bm}
\usepackage{amsfonts}

\begin{document}

\title{Classical Rotons in Cold Atomic Traps}

\author{H. Ter\c{c}as}
\email{htercas@cfif.ist.utl.pt}
\affiliation{CFIF, Instituto Superior T\'{e}cnico, Av. Rovisco Pais 1, 1049-001
Lisboa, Portugal}
\author{J. T. Mendon\c{c}a}
\affiliation{CFIF, Instituto Superior T\'{e}cnico, Av. Rovisco Pais 1, 1049-001
Lisboa, Portugal}
\affiliation{IPFN, Instituto Superior T\'{e}cnico, Av. Rovisco Pais 1, 1049-001
Lisboa, Portugal}
\author{V. Guerra}
\affiliation{IPFN, Instituto Superior T\'{e}cnico, Av. Rovisco Pais 1, 1049-001
Lisboa, Portugal}

\begin{abstract}
We predict the emergence of a roton minimum in the dispersion relation of elementary excitations in cold atomic gases in the presence of diffusive light. In large magneto-topical traps, multiple-scattering of light is responsible for the collective behavior of the system, which is associated to an effective Coulomb-like interaction between the atoms. In optically thick clouds, the re-scattered light undergoes diffusive propagation, which is responsible for a stochastic short-range force acting on the atoms. We show that the dynamical competition between these two forces results on a new polariton mode, which exhibits a roton minimum. Making use of Feynman's formula for the static structure factor, we show that the roton minimum is related to the appearance of long-range order in the system.   
\end{abstract}

\maketitle


Since the first ideas advanced by Landau \cite{landau1, landau2}, the concept of the ``roton minimum" in the dispersion of the collective modes of a certain physical system has played a central role in the description of superfluidity. After the success of the theory in the context of superfluid phases of $^4$He, rotons have received considerable attention since then and have been identified in many different quantum interacting systems. Recently, Cormack et al. \cite{cormack} suggest that rotons may appear in already moderately interacting ultracold Bose gases and Kalman et al. \cite{kalman} have numerically observed their emergence in two-dimensional dipolar bosonic gases. \par
 In fact, the emergence of a roton minimum in the excitation spectrum strongly depends upon the shape of the interacting potential or, equivalently, on how particles are correlated. The correlational origin of the roton minimum has been firstly suggested by Feynman, where the static structure factor $S(k)$ is expressed in terms of the dispersion relation $\omega(k)$ as
 
 \begin{equation}
 \omega(k)=\frac{\hbar k^2}{2mS(k)}.
\label{feynamn}
 \end{equation}
This has an enormous implication on the interpretation of the physical properties of the system in terms of the dispersion relation: the presence of a roton minimum is the signature of strong correlations system. In the limit of large mode softening, i.e., for rotons with zero frequency, the system can develop mechanical instabilities, which can lead to interesting physics phenomena. In Ref. \cite{henkel}, Henkel et al have suggested that the presence of a ``roton zero" is at the origin of crystallization in ultracold Rydberg gases.\par

In this letter, we describe the classical origin of a roton minimum in the excitation spectrum of cold atomic clouds confined in magneto-optical traps (MOTs). Due to competition between long-range interactions between the atoms and the stochastic forces associated to the diffusion of light, atoms in MOTs experience a complex effective interaction, which we find to be associated with a polariton dispersion relation. This polariton mode is result of the dynamical coupling of the density waves with the fluctuations of the light intensity inside the trap. \par

A route for the most intriguing complex behavior in large magneto-optical traps relies exactly in the multiple scattering of light \cite{dalibard, sesko}. Due to the consecutive scattering and re-absorption of photons, the atoms experience a mediated long-range interaction potential similar to Coulomb system ($\sim 1/r$) \cite{walker, pruvost} and the system can therefore be regarded as a one-component trapped plasma. In a series of previous works, we have put in evidence the important consequences of such plasma description of a cold atoms gas \cite{mendonca1, mendonca2}, whereas the formal analogy and the application of plasma physics techniques reveal to be important in the description of driven mechanical instabilities \cite{david, stefano, hennequin, kim, hugo} or even more exciting instability phenomena, like phonon lasing \cite{mendonca3}. Moreover, in such optically thick traps, it is known that the light does not propagate ballistically, rather exhibiting a diffusive behavior \cite{rossum}. In this situation, the energy transport velocity $v_E$, i.e. the velocity that accounts for the propagation of energy by the scattered wave, is smaller than $c$ \cite{albada, tiggelen}. Labeyrie et al. \cite{labeyrie} have experimentally observed that $v_E$ can, indeed, be several orders of magnitude smaller than $c$ in the case of resonant light propagating in traps, already with a moderate optical thickness, thus putting in evidence the phenomenon of slow light. More recently, the diffusive behavior of light has been identified as a source of dynamical instabilities leading to the formation of photon bubbles in magneto-optical traps \cite{mend_bubbles}.\par


In what follows, we consider that the dynamics of cold atoms in MOTs is described by the Vlasov equation 

\begin{equation}
\left(\frac{\partial }{\partial t} + \mathbf{v}\cdot \bm{\nabla}+\frac{1}{m}\sum_i\mathbf{F}_i\cdot \bm{\nabla}_\mathbf{v} \right)f(\mathbf{r},v,t)=0,
\label{vlasov1}
\end{equation}
where $f(\mathbf{r},v,t)$ is the normalized distribution function

\begin{equation}
1=\int d\mathbf{r} \int d\mathbf{v} f(\mathbf{r},v,t). 
\end{equation}
The total force $\sum_i\mathbf{F}_i=\mathbf{F}_T+\mathbf{F}_c$ accounts for both the trapping and cooling forces. There are evidences \cite{pohl2, gattobigio, hugo_eq} that the density profile is approximately constant for large traps (typically with $N\sim 10^9-10^{10}$ atoms), which allows us to consider the system to be homogeneous and thus to neglect the effects of the trap. The collective force can be described by a Poisson equation \cite{pruvost, mendonca1}

\begin{equation}
\bm{\nabla}\cdot \mathbf{F}_c (\mathbf{r},t)=Q_{\mbox{\tiny eff}}\int d\mathbf{v} f(\mathbf{r},v,t),
\label{poisson1}
\end{equation}
The pre-factor in Eq. (\ref{poisson1}) represents an effective charge $Q_{\mbox{\tiny eff}}=\sigma_L(\sigma_R-\sigma_L) I/c$ of the atoms induced by light, where $\sigma_{R}$ and $\sigma_{L}$ represent the scattering and absorption cross sections \cite{walker, dalibard, sesko, pruvost}, and $I$ is the light intensity. For most of the experimental conditions, the scattering cross section is larger than the absorption cross section, i.e. $\sigma_R >\sigma _L$, enforcing the effective charge to be a positive quantity. We have showed that the positiveness of $Q_{\mbox{\tiny eff}}$ is an essential condition for the existence of stable oscillations in the system (see e.g. Ref. \cite{mendonca1}).\par 

We now assume that the diffusive behavior of light inside the trap can be macroscopically described by the diffusion equation

\begin{equation}
\frac{\partial I}{\partial t}-\bm\nabla \cdot \mathcal{D}\bm \nabla I=0.
\label{diff1}
\end{equation}
The diffusion coefficient is determined by $\mathcal{D} = \ell^2 /\tau$ , where the photon mean free pass is $\ell = 1/n \sigma_L$, with $n=n_0\int f d\mathbf{v}$ standing for the atomic density. According to experimental results \cite{labeyrie}, the diffusion time $\tau$ can be considered as independent from the atom density, so the diffusion coefficient explicitly reads 

\begin{equation}
 \mathcal{D}(\mathbf{r},t)=\frac{1}{\sigma_L^2\tau^2 n^2}=\frac{1}{\sigma_L^2\tau^2n_0^2 }\left[\int f(\mathbf{r},\mathbf{v},t) d\mathbf{v}\right]^{-2}.
 \label{diff2}
\end{equation}


We now linearize the Eqs. (\ref{vlasov1}), (\ref{poisson1}) and (\ref{diff2}) by allowing fluctuations around the equilibrium values $f=f_0+\delta f$, $\quad I= I_0 +\delta I$ and $\mathcal{D}=\mathcal{D}_0+\delta \mathcal{D}$, such that

\begin{equation}
 \left(\frac{\partial }{\partial t}+\mathbf{v}\cdot \bm\nabla\right) \delta f +\frac{1}{m}\delta \mathbf{F}_c\cdot \nabla_\mathbf{v} f_0=0
 \label{vlasov_p}
\end{equation}

\begin{equation}
\frac{\partial}{\partial t}\delta I- \mathcal{D}_0\nabla^2 \delta I - \delta\mathcal{D} \nabla^2 I_0=0 
\label{diff_p}
\end{equation}

\begin{equation}
 \bm\nabla\cdot \delta \mathbf{F}_c= Q_{{\mbox{\tiny eff}},0}n_0\int \delta f ~ d\mathbf{v}+Q_{{\mbox{\tiny eff}},0}n_0~\frac{\delta I}{I_0},
 \label{poisson_p}
\end{equation}
where $Q_{{\mbox{\tiny eff}},0}=\sigma_L(\sigma_R-\sigma_L) I_0/c$. Assuming periodic perturbations on both space and time, such that $(\delta f,\delta I, \delta \mathcal{D})\propto \exp(i\mathbf{k}\cdot \mathbf{r}-i\Omega t) $, Eqs. (\ref{diff2}) and (\ref{diff_p}) yield

\begin{equation}
 \delta I=\frac{\beta}{i\Omega -\mathcal{D}_0 k^2}\int d\mathbf{v}\delta f,
 \label{diff_p2}
\end{equation}
where

\begin{equation}
\beta=\frac{2\nabla^2 I_0}{n_0^3\sigma_L^2\tau}=\frac{2\nabla^2 I_0}{n_0}\mathcal{D}_0 
\end{equation}
is the photon inhomogeneity parameter. Combining the latter result with Eq. (\ref{vlasov_p}), we finally obtain the kinetic dispersion relation

\begin{equation}
1=\frac{\omega_p^2}{k^2}\left(1+\frac{\omega_d}{i\Omega-\mathcal{D}_0
k^2}\right)\int \frac{1}{v_z-\Omega/k}\frac{\partial f_0}{\partial
v_x}~d\mathbf{v},
\label{dispersion1}
\end{equation}
where we have considered perturbations parallel to the wave-vector $\mathbf{k}=k e_\mathbf{z}$, for definiteness. Here, we have defined two typical frequencies of the system. The first one is associated with the oscillations of the atoms due to the long-range force, corresponding to an effective plasma frequency \cite{mendonca1}

\begin{equation}
 \omega_p=\sqrt{\frac{Q_{{\mbox{\tiny eff}},0}n_0}{m}}.
\end{equation}
The second important quantity is  the rate at which the photons scatter inside the trap, or simply the {\it diffusion frequency}

\begin{equation}
\omega_d=\frac{\beta n_0}{I_0}=\frac{2\nabla^2 I_0}{I_0}\mathcal{D}_0.
\label{freq_dif}
\end{equation}
We notice that this frequency depends on the scale at which the diffusive processes occur (micro-, meso- or macroscopic), as it depends upon the spatial scale $L$ at which the light intensity varies. We will discuss the macroscopic case below. 

\begin{figure}[ht!]
\centering
\includegraphics[scale=0.7]{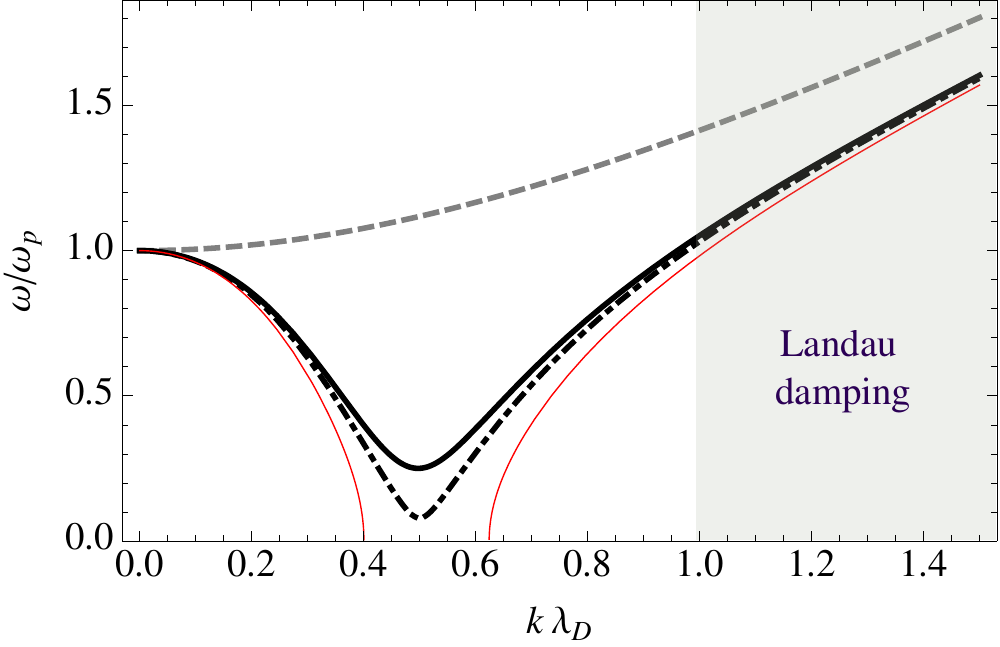}
\includegraphics[scale=0.7]{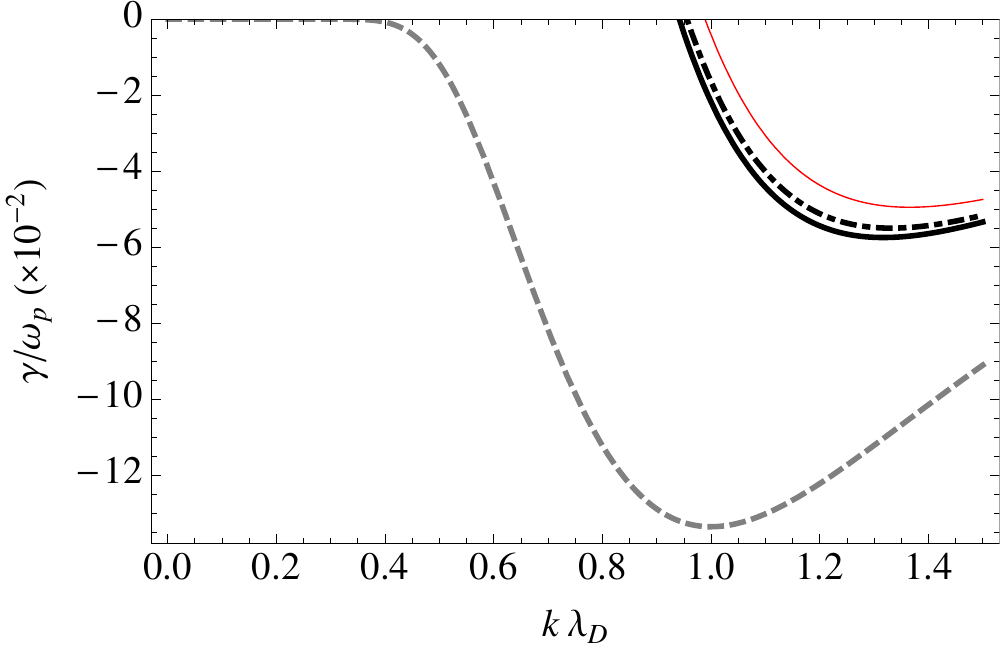}
\caption{(Color online)  Illustration of the real (top panel) and imaginary (bottom panel) parts of the polariton dispersion relation in the macroscopic regime for $\mathcal{D}_0=2.0\lambda_D^2 \omega_p$. We can observe the emergence of a roton minimum for $\omega_d=1.9 \omega_p$ (full black line) and $\omega_d=1.99\omega_p$. The rotons softens the frequency to $\omega(k_{\mbox{\tiny rot}})=0$ at the critical value $\omega_d^{(c)}=2.0\omega_p$. Roton instability is illustrated here for $\omega=2.2\omega_p$ (red line). The short-wavelength oscillations corresponding to $k\lambda_D >1$ are kinematically damped. The usual plasma dispersion relation of ref. \cite{mendonca1} is presented here for comparison (dashed gray line).} 
\label{fig_real}
\end{figure}
The integral in Eq. (\ref{dispersion1}) can be evaluated using the Landau prescription, according to which the full information about the initial conditions is cast if the integration path is set to pass below the pole $\Omega= v_z k$. We split the integral into two parts

\begin{equation}
\begin{array}{r}
\displaystyle{\int \frac{1}{v_z-\Omega/k}\frac{\partial f_0}{\partial v_z} d\mathbf{v}=\mbox{Pr}\int \frac{1}{v_z-\Omega/k}\frac{\partial f_0}{\partial v_z} d\mathbf{v}}\\[15 pt]
\displaystyle{+i\pi \left.\left(\frac{\partial f_0}{\partial v_z}\right) \right \vert_{v_z=\Omega/k}~d\mathbf{v}}, 
\label{principal_value}
\end{array}
\end{equation}
where $\mbox{Pr}$ stands for the Cauchy principal value. Assuming a phase speed $v_{\mbox{\tiny ph}}=\omega/k$ much greater than the width of the distribution, such that $f_0$ and its derivatives get small as $v_z$ gets large, we may expand the denominator in (\ref{principal_value}) which, together with the relation

\begin{equation}
\int \frac{1}{v_z-\Omega/k}\frac{\partial f_0}{\partial v_z} d\mathbf{v}= \int \frac{f_0}{\left(v_z-\Omega/k\right)^2} d\mathbf{v},
\end{equation}
simply yields

\begin{equation}
\mbox{Pr}\int \frac{1}{v_z-\Omega/k}\frac{\partial f_0}{\partial v_z} d\mathbf{v}\simeq
\int f_0\left(1+3\frac{k^2v_z^2}{\Omega^2} \right)dv_x dv_y dv_z.
\end{equation}
Assuming the atomic equilibrium to be described by a Maxwell distribution

\begin{equation}
f_0(v)=\frac{1}{(2\pi v_{th})^{3/2}} e^{-v^2/2 v_{th}^2}, 
\end{equation}
with $v_{\mbox{\tiny th}}=\sqrt{k_B T/m}$ standing for the thermal speed, we may finally write

\begin{equation}
\begin{array}{r}
 \displaystyle{1=\frac{\omega_p^2}{\Omega^2}\left[\left(1+\frac{u_s^2k^2}{\Omega^2} \right)\left(1+\frac{\omega_d}{i\Omega-\mathcal{D}_0 k^2} \right) \right]}\\[15 pt]
 \displaystyle{+i\pi\frac{\omega_p^2\Omega^2}{k^2}\left. \frac{\partial f_0}{\partial v_z}\right \vert_{v_z=\Omega/k}},
 \end{array}
\end{equation}
where we have defined the sound atomic speed $u_s=\sqrt{3}v_{\mbox{\tiny th}}$. Separating the frequency into its real and imaginary parts, $\Omega=\omega +i\gamma$, with $\gamma\ll \omega$, we may finally write 

\begin{equation}
 \omega^2=
\left(\omega_p^2+u_s^2k^2\right)\left(1-\frac{\omega_d\mathcal{D}_0k^2}{\omega_p^2+\mathcal{D}_0^2 k^4}\right)
\label{real_thermal}
\end{equation}
 and
 
 \begin{equation}
  \gamma =
 \frac{\omega_d}{2}\frac{\omega_p^2+u_s^2k^2}{\omega_p^2+\mathcal{D}_0^2
 k^4}-\frac{3}{\sqrt{8\pi}}\frac{1}{k^3\lambda_{D}^3} e^{-3/(2 k^2
 \lambda_{D}^2)},
 \label{imag_thermal}
 \end{equation}
where $\lambda_{D}=u_{S}/\omega_p$ is the effective Debye length. This dispersion relation describes a quasi-particle excitation resulting from the atom-photon coupling, or a polariton. This is the main result of this paper and we now explicitly show that is contains a roton minimum. \par
The diffusive description of light is the result of a macroscopic approximation which is known to hold if absorption takes place at scales much larger than the mean free path $\ell$ \cite{rossum}. This is true provided the following hierarchy for the relevant length scales

\begin{equation}
\lambda\ll \ell \ll a \ll L,
\label{inequality}
\end{equation}
where $\lambda$ is the light wavelength and $a$ is the size of the system. According to typical experimental conditions \cite{labeyrie}, the mean-free path is found to value $\ell\sim 300$ $\mu$m and the diffusion coefficient $\mathcal{D}_0\simeq 0.66$ m$^{2}$s$^{-1}$. Based on our previous estimates \cite{mendonca1}, the effective plasma frequency and Debye length respectively value $\omega_p \sim 2\pi \times 100$ Hz  and $\lambda_D\sim 100$ $\mu$m. Therefore, provided the identification $L=2\nabla^2 I_0/ I_0$ in (\ref{freq_dif}) and using the inequalities in Eq. (\ref{inequality}) with $a\sim 1$ mm \cite{labeyrie}, the diffusive approximation is valid if the diffusion and plasma frequencies are of the same order, i.e. $\omega_d\sim \omega_p$. This is attained if $\ell_d\equiv \sqrt{D_0/\omega_p}\sim 1$ cm is of the same order of the intensity variation length $L$, which is a reasonable condition for typical experimental scenarios. Nevertheless, by varying the number of atoms in the trap it is possible to control the optical thickness of the system and, therefore, tune the value of the diffusion coefficient $\mathcal{D}_0$, which makes our present estimates even more flexible. In extremis, it may be also possible to attain different diffusion regimes, but this situation is out of the scope of the present paper. In Fig. (\ref{fig_real}), it is shown that a roton minimum emerges in the dispersion relation (\ref{real_thermal}) in the diffusive regime. As the values of $\omega_d$ increase (i.e., for stronger diffusion), the frequency decreases around the roton wavenumber $k_{\mbox{\tiny rot}}\sim \lambda_D/\ell_d^2$. This feature is often referred to as mode softening. At the critical value $\omega_d^{(c)}=2\omega_p$, the mode softens towards zero, which is a clear manifestation of a roton instability mechanism. For $\omega_d>\omega_d^{(c)}$, the system enters a crystallization phase. This mechanism has been recently discussed in the literature as it can lead to the formation of supersolids \cite{henkel}. An important remark is related to the Landau damping at short wavelengths. Modes in the region $k\lambda_D \gtrsim 1$ undergo a kinematic damping. Fortunately, rotons are possible to be excited at longer wavelengths ($k_{\mbox{\tiny rot}} \lambda_D<1$), thus avoiding the Landau damping mechanism.  Moreover, the onset of diffusion tends to decrease the damping rate (see Fig. (\ref{fig_real})). This nourishes hope for rotons to be experimentally observable.\par 

A remarkable feature of the polariton spectrum in (\ref{real_thermal}) is that it exhibits a roton minimum in a three dimensional system even in the absence of strong interactions. It is clear from the application of Landau's criterion that the present spectrum does not correspond to that of a superfluid, as the mode is gapped at the origin. This is a consequence of the long-range nature of the interaction between nearby atoms, and the quasi-particle mass is proportional to $\omega_p$. This therefore corresponds to an important example where rotons are not intrinsically connected to Goldstone modes.\par
\begin{figure}[t!]
\includegraphics[scale=0.75]{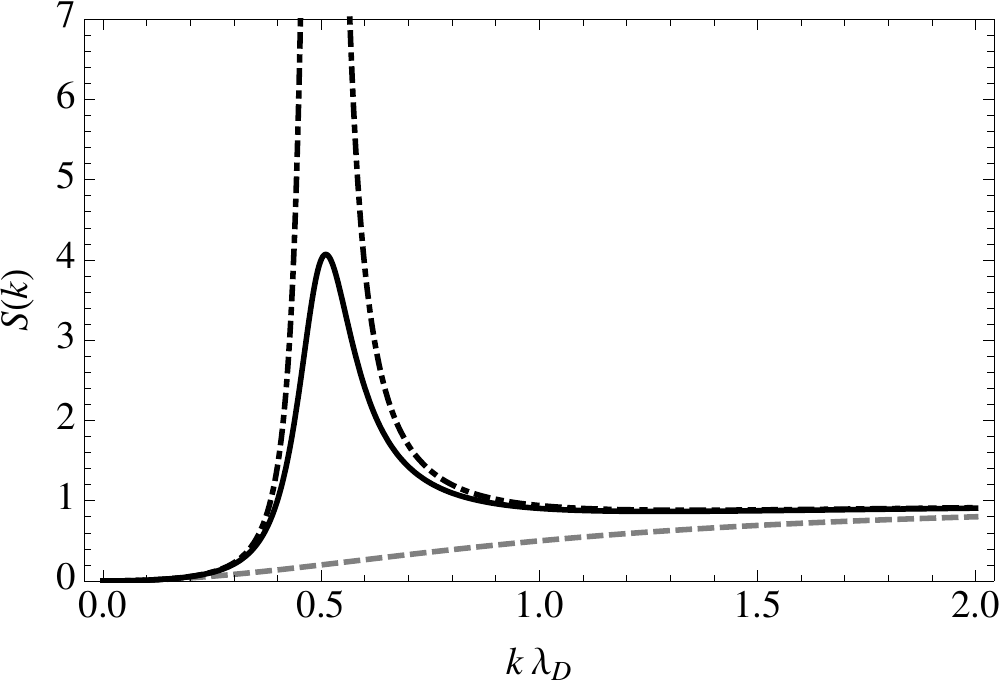}
\caption{Static structure factor $S(k)$ depicted for $\mathcal{D}_0=2.0\lambda_D^2 \omega_p$. $\omega_d=0$ (gray dashed line), $\omega_d=1.9 \omega_p$ (black full line) and $\omega_d=1.99\omega_p$ (black dashed line). A pick emerges in the static structure factor around $k_{\mbox{\tiny rot}}=\lambda_D/\ell_d^2=0.5\lambda_D^{-1}$.}
\label{fig_static}
\end{figure}

Another important property of the classical rotons described above is that they carry useful information about the long-range correlation of the system. By using the extension of Feynman's formula (\ref{feynamn}) to finite temperature systems \cite{feynman2}, the static structure factor $S(k)$ is given by

\begin{equation}
S(k)=\frac{\hbar k^2}{2m\omega(k)}\coth\left(\frac{\hbar\omega(k)}{2k_BT}\right)\simeq \frac{u_s^2 k^2}{\omega(k)^2}, 
\label{SSF}
\end{equation}
where we have used the non-degeneracy condition $k_BT\gg \hbar\omega$. In this limit, the structure factor corresponds to that obtained based on a hydrodynamic treatment \cite{wang}. In Fig. (\ref{fig_static}), we illustrate the behavior of $S(k)$ for the same parameters of Fig. (\ref{fig_real}).

\begin{figure}[hb!]
\includegraphics[scale=0.75]{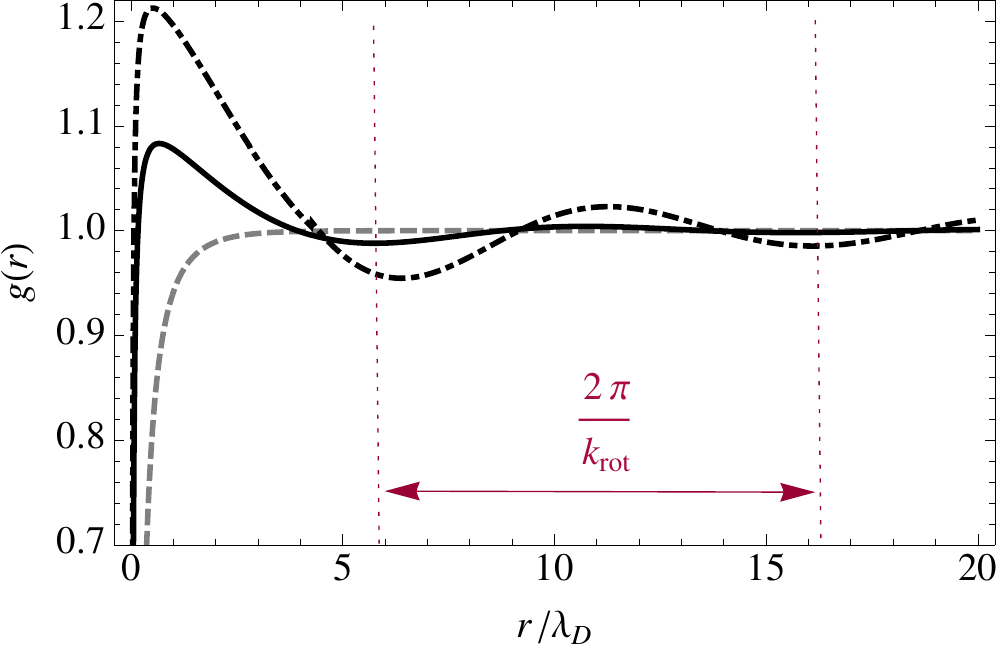}
\caption{Two-point correlation function $g(r)$ depicted for $\mathcal{D}_0=2.0\lambda_D^2 \omega_p$. $\omega_d=0$ (gray dashed line), $\omega_d=1.9 \omega_p$ (black full line) and $\omega_d=1.99\omega_p$ (black dashed line). The correlation function oscillates with the period a of $2\pi/k_{\mbox{\tiny rot}}\simeq 4\pi \lambda_D^{-1}$.}
\label{fig_corr}
\end{figure}

The static two-point correlation function $g(r)=\langle n(r) n(0)\rangle /\langle n(r)\rangle \langle n(0)\rangle \simeq \langle n(r) n(r')\rangle /n_0^2$ can then be easily calculated provided the relation $g(r)=1+\mathcal{F}^{-1}\left[S(k)-1\right]$ \cite{ashraf}, which after the integrating out the angular variables simply reads 

\begin{equation}
 g(r)=1+\frac{1}{\pi^2}\int_0^\infty \frac{k\sin(kr)}{r}\left[S(k)-1\right].
\label{CF}
\end{equation}
As it can be observed in Fig. (\ref{fig_corr}), the appearance of a minimum in the excitation spectrum (\ref{real_thermal}) is associated with the occurrence of long-range correlation in the system. By inspection, one founds that the correlation function oscillates with the period $T=2\pi/k_{\mbox{\tiny rot}}$. This feature can be qualitatively understood in the context of Percus-Yevick theory \cite{percus,wertheim}, where the correlation function is approximated by $g(r)^{\mbox{\tiny PV}}\simeq 1+c_0 r^{-1}\cos(k_0+\delta_0)e^{-\kappa_0 r}$, where $c_0$ is a constant and $z_0=\kappa_0+ik_0$ is the pole of the function $S(k)-1$. We remark, however, that the PY theory was originally developed for hard-sphere potentials, and therefore does not describe systems with long-range interactions. For that reason, we have not used it to compute $g(r)$. 

In conclusion, we have derived the dispersion relation for the atom-photon polariton in large magneto-optical traps in the presence of diffusive light. We have explicitly computed the excitation spectrum for the particular case of a thermal atomic distribution, revealing the emergence of a roton minimum for a set of parameters compatible with current experimental conditions. We have also shown that the increase of the light diffusivity lead to mode softening around the roton minimum, which may eventually drive the system into a roton instability situation. Using the relation between the static structure factor and the dispersion relation, we have explicitly demonstrated that the roton minimum is related to the emergence of long-range correlations in the system.

\par
This work was partially supported by Funda\c c\~ao para a Ci\^encia e Tecnologia (FCT-Portugal), through the grant number SFRH/BD/37452/2007.


\begin{thebibliography}{10}
\bibitem{note1} We here consider that the total incident light intensity $I$ results from the six laser beams used in standard MOT configurations.
\end{thebibliography}
\end{document}